\documentclass[aps,onecolumn,11pt,floatfix,altaffilletter,superscriptaddress,
               preprintnumbers,tightenlines,showpacs,showkeys,notitlepage,nofootinbib]{revtex4-2}

\usepackage[colorlinks=true,citecolor=blue,linkcolor=blue]{hyperref}
\usepackage[normalem]{ulem}
\usepackage{amsmath,amssymb}
\usepackage{mathtools}
\usepackage{verbatim}
\usepackage{titlesec}               
\usepackage{epsfig}
\usepackage{graphicx}               
\usepackage{url}
\usepackage{color}
\usepackage{multirow}
\usepackage{placeins}
\usepackage[dvipsnames]{xcolor}
\usepackage{epstopdf}
\usepackage{siunitx}
\usepackage{fontawesome}
\usepackage{enumitem}
\usepackage[capitalize]{cleveref}
\usepackage{lipsum}
\usepackage{gensymb}
\usepackage{aas_macros}             
\usepackage{booktabs}               
\usepackage{xspace}                 
\usepackage{pifont}
\usepackage{marvosym }
\usepackage{orcidlink}
\allowdisplaybreaks

\makeatletter

\renewcommand{\p@subsection}{}
\renewcommand{\p@subsubsection}{}
\makeatother

\titleformat*{\section}{\centering\bfseries\uppercase}
\titlelabel{\thetitle\quad}
\titleformat*{\paragraph}{\bfseries}
\titlespacing*{\paragraph}{0pt}{3.25ex plus 1ex minus .2ex}{1em}

\makeatletter
\def\l@subsubsection#1#2{}
\makeatother

\begin{document}

\newcommand{\prisma}{
    PRISMA${}^+$ Cluster of Excellence \& MITP,
    Johannes Gutenberg University, 55099 Mainz, Germany}

\title{A LENS on DUNE-PRISM: \\ Characterizing a Neutrino Beam with Off-Axis Measurements}

\author{Julia Gehrlein \orcidlink{0000-0002-1235-0505}\,}
\email{julia.gehrlein@colostate.edu}
\affiliation{Physics Department, Colorado State University, Fort Collins, CO 80523, USA}

\author{Joachim Kopp \orcidlink{0000-0003-0600-4996}\,}
\email{jkopp@uni-mainz.de}
\affiliation{\prisma}
\affiliation{Theoretical Physics Department, CERN,
             1 Esplanade des Particules, 1211 Geneva 23, Switzerland}

\author{Margot MacMahon \orcidlink{0000-0003-2604-3415}\,}
\email{margot.macmahon.21@ucl.ac.uk}
\affiliation{Department of Physics and Astronomy, University College London, London WC1E 6BT, United Kingdom}

\author{George A.\ Parker \orcidlink{0009-0000-1836-8696}\,}
\email{geparker@uni-mainz.de}
\affiliation{\prisma}

\date{\today}
\preprint{MITP-25-065}

\begin{abstract}
\noindent
Upcoming precision long-baseline neutrino oscillation experiments will be severely limited by the large systematic uncertainties associated with neutrino flux predictions and neutrino-nucleus cross sections. A promising remedy is the PRISM (Precision Reaction Independent Spectrum Measurement) technique, whereby the near detector measures the neutrino spectrum at different angles with respect to the beam axis. These measurements are then linearly combined into a prediction of the oscillated neutrino flux at the far detector. This prediction is data-driven, but still dependent on some theoretical knowledge about the neutrino flux. In this paper, we study to what extent off-axis measurements themselves can be used to directly constrain neutrino flux models. In particular, we use them to extract separately the fluxes and spectra of different meson species in the beam. We call this measurement LENS (\textbf{L}ateral \textbf{E}xtraction of \textbf{N}eutrino \textbf{S}pectra). Second, we demonstrate how the thus improved flux model helps to further constrain the far detector flux prediction, thereby ultimately improving oscillation measurements.
\end{abstract}

\maketitle

\tableofcontents
\clearpage

\section{Introduction}
\label{sec:intro}

Neutrino physics has long been a game of small numbers: given the feebleness of neutrino interactions, the field has traditionally been constrained by the small event numbers that could be achieved. But no more. Powerful accelerator-driven neutrino sources, enormous detectors, and sophisticated analysis techniques are turning neutrino physics into a precision science, so that statistical errors become less and less of a concern while systematic uncertainties come into sharp focus. The systematic error budget of upcoming long-baseline oscillation experiments like DUNE \cite{DUNE:2020ypp} and Hyper-Kamiokande \cite{Hyper-Kamiokande:2018ofw,Hyper-Kamiokande:2025fci} reflects in particular our limited understanding of neutrino fluxes and spectra from accelerator sources, and it mirrors our imperfect knowledge of neutrino-nucleus interaction cross sections at GeV energies.

Accelerator neutrino beams are created from the decays of mesons (primarily pions and kaons) produced when a high-energy, high-intensity proton beam strikes a target. Uncertainties in hadron production translate directly into uncertainties in the neutrino flux. Accordingly, dedicated hadroproduction experiments have been critical in sharpening flux predictions. The NA61/SHINE \cite{NA49-future:2006qne, NA61:2014lfx,NA61SHINE:2025aey} experiment at CERN, for instance, has measured secondary particle yields for various target materials and geometries. 
The EMPHATIC \cite{EMPHATIC:2019xmc} collaboration proposes to collect new hadron
production data to reduce neutrino flux uncertainties using the Fermilab Test Beam Facility. These data, along with results from earlier experiments (e.g.\ HARP at CERN \cite{HARP:2007wvs} and MIPP at Fermilab \cite{MIPP:2014shj}) provide crucial calibration of hadronic interaction models, thereby reducing the neutrino flux uncertainty entering oscillation analyses to the few-percent level. 
Looking forward, there are ambitious ideas for a flux spectrometer in the DUNE beamline \cite{fluxspectro}; for a monitored neutrino beam at CERN (ENUBET), in which the meson decays producing the beam are monitored in real time in an instrumented decay tunnel \cite{ENUBET:2023hgu}; and even a tagged neutrino beam NuTag), in which each detected neutrino could be matched to its parent meson decay and the full kinematics of that decay could be reconstructed \cite{Baratto-Roldan:2024bxk}. ENUBET and NuTag have recently merged into the nuSCOPE project \cite{Acerbi:2025wzo}.

This class of experiments will improve neutrino interaction cross-section measurements with sophisticated in-situ
monitoring of the neutrino flux.
For now, however, neutrino flux uncertainties remain a major concern for precision neutrino experiments.

To limit the extent to which flux and cross-section uncertainties can obscure the sought-after oscillation patterns, modern long-baseline neutrino oscillation experiments employ a sophisticated suite of near detectors (placed close to the source) to characterise the unoscillated neutrino flux and measure the cross sections. A novel feature in DUNE and Hyper-Kamiokande will be the ability to move some of these detectors transverse to the beam axis, a concept called ``PRISM'' (``Precision Reaction Independent Spectrum Measurement'') \cite{nuPRISM:2014mzw}. The usefulness of PRISM stems from the fact that the neutrino flux and spectrum change as a function of the off-axis angle $\theta_\text{oa}$, while the cross sections do not. Therefore, PRISM disentangles flux and cross-section uncertainties.

Before proceeding, it is useful to clarify some terminology:
\begin{itemize}[leftmargin=*]
    \item\textbf{PRISM concept:} The use of a movable detector, multiple detectors, or different parts of the same detector to view neutrino spectra at different angles off the beam axis \cite{nuPRISM:2014mzw}.
    \item\textbf{PRISM prediction:} A specific application of PRISM is a data-driven prediction of the oscillated flux at the far detector. More specifically, the far detector spectrum is written as a linear combination of near detector spectra taken at different $\theta_\text{oa}$ \cite{nuPRISM:2014mzw, DUNE:2021tad, Hasnip:2023ygr, Hasnip:2025gyi, DUNE:2025lvs} to improve oscillation and cross-section measurements.\footnote{The method of predicting far detector fluxes should have its own name, and we humbly suggest ``HELICOPTER'' (``HElpful LInear COmbination of PRISM TEmplate Rates'').}
    \item\textbf{PRISM-LENS} (Lateral Extraction of Neutrino Spectra): Another application of the PRISM concept, where the different off-axis spectra are used to constrain the flux model, which is the focus of this work. This flux model can then be employed in the PRISM prediction for the far detector spectrum.
\end{itemize}
Specific realizations of the PRISM concept are foreseen in both DUNE and Hyper-Kamiokande. In DUNE, part of the near detector complex will be able to move laterally off the beam axis by up to \SI{36}{m}, which at a distance of \SI{574}{m} corresponds to off-axis angles up to 3.6~degrees. Hyper-Kamiokande is planning to deploy a movable ``Intermediate Water Cherenkov Detector'' (IWCD) about \SI{1}{km} from the neutrino source, with the capability to move vertically in order to probe off-axis angles from $1^\circ$ to $4^\circ$ \cite{nuPRISM:2014mzw, Scott:2016kdg, Mondal:2024agd}. Also the Short-Baseline Near Detector (SBND) at Fermilab’s Booster Neutrino Beam, though immovable, can explore the transverse profile of the neutrino beam thanks to its height and width of about \SI{4}{m} \cite{tuttoaps, tuttocewg,Abratenko:2025nfv}. Combined with a slight (\SI{74}{cm}) displacement of the detector centre with respect to the beam axis, this translates into off-axis angles up to $1.4^\circ$ \cite{SBND:2025lha}. 

However, even with innovative data-driven methods like those enabled by the PRISM concept, precise a priori knowledge of the neutrino flux remains important. 
A major uncertainty at DUNE will come from the focusing of the mesons. These uncertainties  affect the dependence of the flux on the off-axis angle, something the PRISM concept is not robust against.

In this context, we suggest PRISM-LENS (from now on, just LENS), which refers to first carrying out a fit to near detector data alone to refine the flux model. We will see that using a refined flux model improves the accuracy of the PRISM prediction, and will demonstrate how this impacts oscillation measurements. The LENS approach complements the external hadroproduction measurements discussed above. We note that in \cite{Karpova:2025lvb} a different approach to constrain neutrino flux parameters using off-axis detector positions has been put forward.

While this work is mostly about DUNE, we emphasise that our conclusions apply analogously also to Hyper-Kamiokande with the movable IWCD near detector and SBND due to its width and proximity to the target. Focusing on DUNE has simple practical reasons: beam fluxes and detector response functions for DUNE are public \cite{DUNE:2021cuw}, allowing us to simulate the experiment more reliably.

The plan of the paper is as follows: in \cref{sec:prism}, we will review the technicalities of the PRISM prediction. In \cref{sec:lens}, we will then quantify the accuracy with which DUNE-PRISM can characterise the neutrino flux. Finally, in \cref{sec:osc}, we investigate how flux uncertainties with and without PRISM constraints impact oscillation measurements. We conclude in \cref{sec:conclusions}.

\section{The PRISM Prediction of the Oscillated Far Detector Flux}
\label{sec:prism}

\begin{figure}
    \centering
    \includegraphics[width=\textwidth]{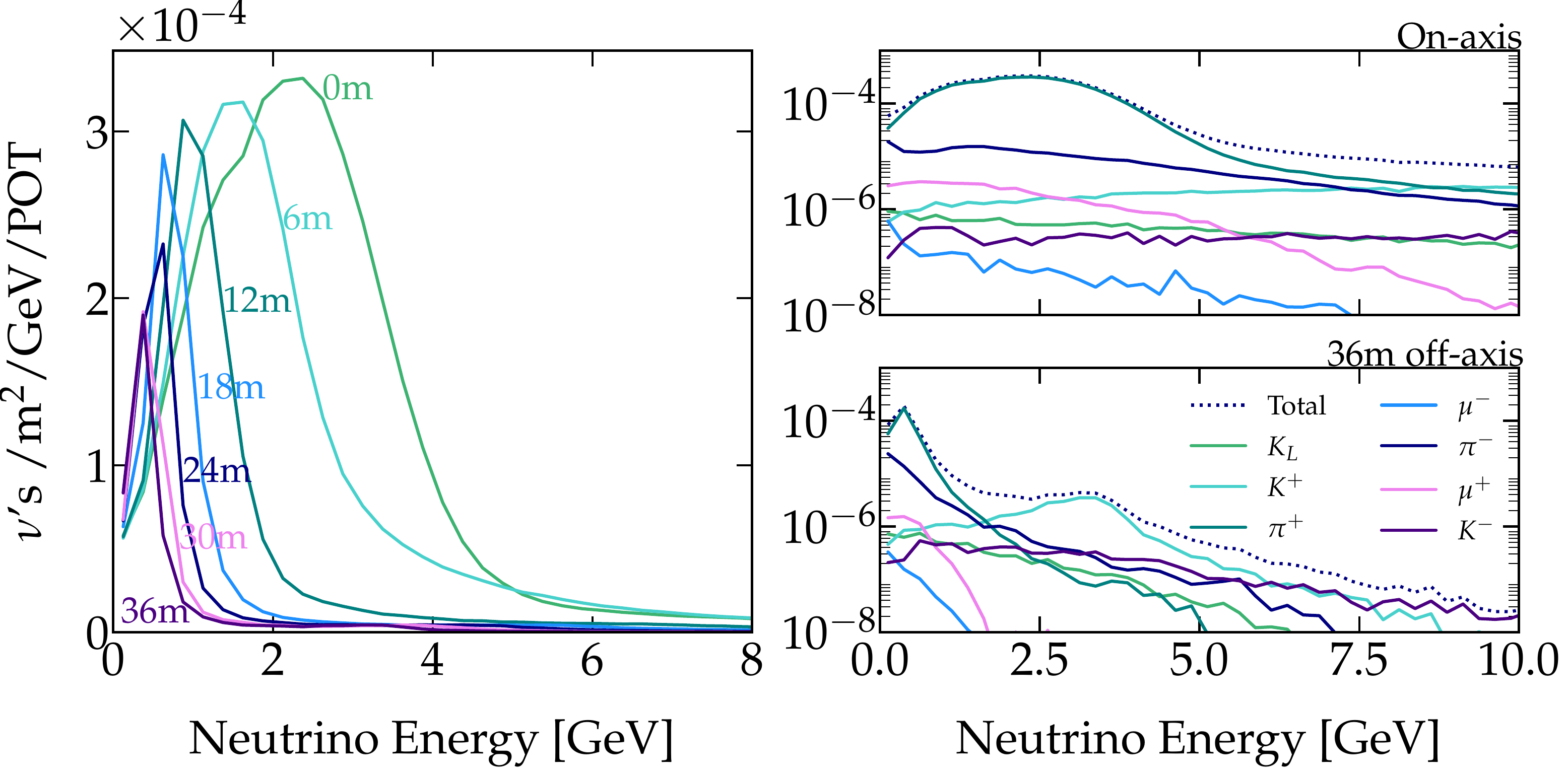}
    \caption{Neutrino flux in the DUNE/LBNF neutrino-dominated beam at different off-axis angles (left). For the on-axis ($0^\circ$) and maximally off-axis ($3.59^\circ$) positions, we also show separately the different flux components corresponding to different parent particles (right).}
    \label{fig:fluxes}
\end{figure}

The PRISM concept exploits the fact that the peak of the neutrino spectrum moves to lower energies away from the beam axis, see \cref{fig:fluxes} \cite{nuPRISM:2014mzw, DUNE:2021tad, Hasnip:2025gyi}. This can be understood from simple kinematics: higher energy neutrinos originate from parent particles with larger Lorentz boosts, implying more forward decays. For the same reason, the relative contributions of neutrinos from different parent particle species change with the off-axis angle $\theta_\text{oa}$. For instance, pion decays dominate the on-axis flux, while their relative importance compared to kaons is reduced at $\theta_\text{oa} > 0$. These changes to the flux as a function of $\theta_\text{oa}$ are reasonably well understood, as decay kinematics are well understood. What is not well understood are (1) the absolute normalizations or shapes of the different flux components and (2) the neutrino cross-sections.

The method that has been proposed for predicting oscillated far detector spectra based on PRISM measurement (PRISM prediction) combines near detector spectra collected at different $\theta_\text{oa}$, $\phi_\text{ND}(E; \theta_\text{oa})$, into a prediction of the (oscillated) far detector flux, $\phi_\text{FD}(E; \vec\Theta)$. The prediction obtained in this way, which we will call the PRISM prediction, is robust with respect to cross-section uncertainties. The idea is to write
\begin{align}
    \phi_\text{FD}(E_i; \vec\Theta) = \sum_j c_j(\Theta) \, \phi_\text{ND}(E_i; \theta_{\text{oa},j}) \,,
    \label{eq:PRISM-1}
\end{align}
where $\vec\Theta = (\theta_{12}, \theta_{13}, \theta_{23}, \delta_\text{CP}, \Delta m_{21}^2, \Delta m_{31}^2)$ denotes the six oscillation parameters, the subscript $i$ denotes the different energy bins, and the index $j$ labels different off-axis angles at which the spectrum has been measured by the near detectors. In the following, we will use the shorthand notation $\phi_{\text{ND},ij} \equiv \phi_\text{ND}(E_i; \theta_{\text{oa},j})$ and $\phi_{\text{FD},i} \equiv \phi_\text{FD}(E_i; \vec\Theta)$.
The coefficients of the linear combination, $c_j(\Theta)$, are determined on the basis of simulated near- and far-detector fluxes $\phi_{\text{ND},ij}^\text{MC}$, $\phi_{\text{FD},i}^\text{MC}$ as
\begin{align}
    c_j = \sum_i \big[ (\phi_{\text{ND}}^\text{MC})^{-1} \big]_{ji} \phi_{\text{FD},i}^\text{MC} \,.
    \label{eq:PRISM-coefficients}
\end{align}
This needs to be done separately for each set of oscillation parameters. The $c_j$ are then applied to the \emph{observed} near detector spectra\footnote{Note the observed fluxes, $\phi^{\text{obs}}$, have to be extracted from the observed number of events $N^{\text{obs}}$ and the cross section.} $\phi_{\text{ND},ij}^\text{obs}$ to obtain a far detector prediction, $\phi_{\text{FD},i}^\text{PRISM}$ with significantly reduced systematic uncertainties compared to $\phi_{\text{FD},i}^\text{MC}$:
\begin{align}
    \phi_{\text{FD},i}^\text{PRISM} = \sum_j c_j \, \phi_{\text{ND},ij}^\text{obs} \,.
    \label{eq:PRISM-prediction}
\end{align}
This prediction is finally compared to the far detector data using a standard maximum likelihood analysis.

In practice, either a few extra steps are needed, in particular unfolding of the detector response and subtraction of backgrounds in $\phi_{\text{ND},ij}^\text{obs}$, and addition of detector response and backgrounds in $\phi_{\text{FD},i}^\text{PRISM}$ before the latter can be compared to data. In addition, the near detectors measure mainly $\nu_\mu$, while in the far detector, the (oscillated) $\nu_e$ flux is of greatest interest for measurements such as the search for leptonic CP violation, the determination of the neutrino mass ordering, etc. Therefore, the corresponding cross section differences (due to the different masses of $e$ and $\mu$) should be taken into account. Finally, care must be taken in evaluating \cref{eq:PRISM-coefficients}, which constitutes an ill-posed linear inverse problem. Typical solutions lead to large variations between adjacent $c_j$ and fine-tuned cancellations in \cref{eq:PRISM-prediction}. Statistical fluctuations in $\phi_{\text{ND},ij}^\text{obs}$ are then amplified, limiting the precision of the method. As a remedy, the DUNE collaboration have explored the Tikhonov regularisation procedure \cite{Phillips:1962ofa, Willoughby1979}, which adds a penalty term to the least squares method, aiming to reduce statistical uncertainty in the PRISM prediction \cite{Hasnip:2023ygr, DUNE:2025lvs}. In this exploratory work, we will not consider regularisation. This will isolate the effect of uncertainty reduction via the LENS procedure. However, in a full experimental analysis, we would expect LENS to be implemented alongside a regularisation scheme.

The key feature of the PRISM prediction is that $\phi_{\text{FD},i}^\text{PRISM}$ is largely insensitive to errors on the neutrino cross sections and total neutrino fluxes as these uncertainties affect the near and far detectors in the same way. Moreover, this could reduce the extent to which the detector-response correction is coupled to assumptions about the neutrino-interaction model and the differences in the event spectra  between the near and far detectors as long as these effects are correctly modelled in the Monte Carlo simulation. 

The method is, however, vulnerable to mismodelling of the \emph{relative} normalisation of different flux components. Such mismodelling leads to differences between the true and assumed $\theta_\text{oa}$ dependence of the fluxes, thereby biasing the extraction of the $c_j$ coefficients in \cref{eq:PRISM-coefficients}. Our goal in the following will be to quantify the accuracy with which a movable near detector itself can characterise the different flux components, and how the residual uncertainties affect oscillation measurements.

\section{Lateral Extraction of Neutrino Spectra (LENS)}
\label{sec:lens}

Flux and spectrum measurements at different off-axis angles can be used to constrain neutrino flux models and to adapt them to the actual operating conditions of the beam. We refer to such measurements as LENS (Lateral Extraction of Neutrino Spectra). LENS leads to improved predictions for $\phi_{\text{ND}}^\text{MC}$ and $\phi_{\text{FD}}^\text{MC}$ in \cref{eq:PRISM-coefficients} and thereby to more reliable values for the coefficients $c_j$ and a better prediction for the oscillated far detector flux $\phi_{\text{FD}}^\text{PRISM}$ in \cref{eq:PRISM-prediction}.

\subsection{Theoretical Modelling of the DUNE Neutrino Flux}
\label{sec:flux-modelling}

The first step is to construct a model for the neutrino flux as a function of $E$ and $\theta_\text{oa}$, depending on a set of parameters which can then be determined from the on-axis and off-axis near detector data. The flux model should provide enough freedom to reproduce any imaginable bias in the true flux. At the same time, the number of free parameters should not be too large, otherwise the model loses predictivity. Here, we separate the neutrino flux into five separate categories according the neutrino's parent particle, $K_L$, $K^+$, $K^-$,  $\pi^+$, and $\pi^-$. The main free parameters of the model are the normalizations of these components. We do not treat neutrinos from muon decay as a separate category, given that their flux is small, and the normalization is correlated with the pion and kaon fluxes from which muons originate. Clearly, this flux model, while sufficient for our exploratory study, is not meant to be final. Our model does, however, capture the largest sources of uncertainty, i.e.\ the ones that will become important first as oscillation measurements move from the statistics-dominated regime to the systematics-dominated regime.

We extract the nominal neutrino spectra from $K_L$, $K^+$, $K^-$, $\pi^+$, and $\pi^-$ decay from the ROOT Ntuple files provided by the DUNE collaboration \cite{fieldsfluxes}, representing the output of DUNE's beamline Monte Carlo simulation. For each simulated neutrino, we look up the direct parent particle and classify the neutrinos accordingly into one of our five categories. Given the limited size of the DUNE Ntuples, this procedure leads to insufficient Monte Carlo statistics. We therefore re-decay each parent meson 200 times using the \texttt{phasespace} package \cite{puig_eschle_phasespace-2019} to generate additional random kinematic configurations without affecting the total exposure. To determine the correct geometric weight factor for each neutrino, we propagate the flux through the near detector and determine the distance it travels inside using the \texttt{trimesh} package \cite{trimesh}. We focus here on the ND-LAr (liquid argon) detector, which we describe as a cuboid \SI{7}{m} wide (transverse to the beam axis), \SI{3}{m} high, and \SI{5}{m} long (along the beam axis). It is located \SI{574}{m} from the target, and its fiducial mass is \SI{67}{ton} \cite{DUNE:2021tad}.

To obtain predictions for the event spectra observed at the near detector, we fold the neutrino fluxes with the cross sections, $\sigma_\alpha(E)$, for $\nu_\alpha$--argon scattering (calculated in GENIE V3\_04\_00, configuration AR23\_20i\_00\_000 \cite{Andreopoulos:2009rq}) and with the detector response function $R_\alpha(E_\text{reco}, E)$ \cite{DUNE:2021cuw}.

We finally also add the DUNE backgrounds, $B_{\alpha,ij}$, \cite{DUNE:2021cuw} to obtain the event rate $N_{\alpha,ij}$ in the $i$-th energy bin and $j$-th angular bin,
\begin{align}
    N_{\alpha,ij} = \int \! dE_\nu \, (\phi_{\text{ND},\alpha,ij}+B_{\alpha,ij}) \, \sigma_\alpha(E_i) \, R_\alpha(E_{\text{reco},i}, E_\nu) 
\end{align}
The dependence of the backgrounds on the normalization of the neutrino flux components have been taken into account as well.
We include Gaussian energy smearing, based on the DUNE GLoBES implementation \cite{DUNE:2021cuw}.

\subsection{Characterising the Neutrino Flux with LENS}
\label{sec:flux-constraints}

We now show how off-axis measurements can help to characterise the neutrino beam. In particular, we estimate the accuracy with which the individual components of the neutrino flux (corresponding to different parent particles in our flux model) can be reconstructed. This is possible only because the fluxes depend on $\theta_\text{oa}$ in a well-known way, whereas the cross sections do not depend on the angle. This allows us to disentangle flux and cross-section uncertainties.

We write the flux for each flavour $\alpha$ as the sum of five contributions, corresponding to neutrinos with different parent particles:
\begin{align}
    \phi_{\text{ND},\alpha}(E; \theta_{\text{oa}}) = \sum_p r_p \, \phi_{\text{ND},\alpha}^{(p)}(E; \theta_{\text{oa}}) \,,
    \label{eq:flux-superposition}
\end{align}
where the superscript $p = K_L, K^+, K^-, \pi^+, \pi^-$ denotes the parent particles. The coefficients $r_p$ are our fit parameters. For the nominal flux prediction, $r_p = 1$ for all $p$. Note that the $r_p$ depend neither on $\theta_\text{oa}$ nor on $E$. Additionally, the shape of the meson flux would be affected by, for example, variations in the horn current. In the following, we account for such changes by implementing a bin-by-bin systematic uncertainty.

For our sensitivity study, we perform a frequentist maximum-likelihood fit to an Asimov data set of simulated off-axis spectra at $7$ different off-axis angles between $0^\circ$ (on-axis) and $3.59^\circ$. We include separate spectra for $\nu_\mu$, $\bar\nu_\mu$, $\nu_e$, $\bar\nu_e$. We base this ``data'' on the nominal DUNE fluxes (QGSP\_BERT model, version glbne-v3r5p10-22-gf42572f) from Ref.~\cite{fieldsfluxes}. In the fit, we vary the $r_p$ coefficients as well as several nuisance parameters related to the neutrino cross sections. More precisely, we write the neutrino cross section as 
\begin{align}
    \sigma_{\nu_\alpha} &= \sum_{k=\text{QE},\text{MEC},\text{RES},\text{DIS}} a_k \sigma_{\nu_\alpha,k}\left(\frac{E}{E_0}\right)^{\gamma_k},
    &\text{$\alpha = e, \mu$} ,
\end{align}
separating it into contributions from quasi-elastic scattering (QE), processes with meson exchange currents such as two-particle--two-hole processes (MEC), QCD resonance production (RES), and deep-inelastic scattering (DIS). The nuisance parameters $a_k$ describe normalization uncertainties for these partial cross sections, while the $\gamma_k$ correspond to spectral tilts. The nominal cross sections correspond to $a_k = 1$, $\gamma_k = 0$. We include pull terms for the nuisance parameters, constraining them to within the ranges given in \cref{tab:sys_uncertainties}. The fractional systematic uncertainties  on $a_k,~\gamma_k$ are chosen   such that when varying them within the quoted ranges,  the  GENIE predictions for the cross-section components encompasses the NuWro predictions on these cross section components.
This treatment of the cross-section uncertainties is conservative and it additionally includes a 1.5\% uncertainty on detector response and efficiency from Ref.~\cite{DUNE:2020ypp}. 

\begin{table}
    \centering
    \addtolength{\tabcolsep}{4pt}
    \begin{tabular}{cc@{\ \ }cc@{\ \ }c}
        \toprule
        \textbf{Parameter}
           & \multicolumn{4}{c}{\textbf{Systematic uncertainty}} \\
           & QE & MEC & RES & DIS\\
        \cmidrule(lr){2-5} 
                  $a$      & $7\%$ & $8\%$ & $8\%$ & $11\%$ \\
          $\gamma$ & $4\%$ & $10\%$ & $10\%$ & $3\%$ \\
        \bottomrule
        \end{tabular}
    \addtolength{\tabcolsep}{-4pt}
    \caption{Fractional systematic uncertainties from our neutrino interaction cross-section model. 
    These include a $1.5\%$ uncertainty on detector response and efficiency from Ref.~\cite{DUNE:2020ypp}. }
    \label{tab:sys_uncertainties}
\end{table}

Our likelihood function then has the standard form
\begin{align}
    -2 \log \mathcal{L}(\vec{r})
        = \min_{\vec{a}} \Bigg[ \sum_{\substack{i=\text{energies} \\ j=\text{angles} \\ \alpha=\text{flavors}}}
              \Bigg( \frac{N^\text{obs}_{\alpha,ij} - N^\text{pred}_{\alpha,ij}(\vec{r}; \vec{a})}
                          {N^\text{pred}_{\alpha,ij}(\vec{r}; \vec{a})+(0.01 N^\text{pred}_{\alpha,ij}(\vec{r}; \vec{a}))^2} \bigg)^2
            + \sum_k \frac{(a_k - 1)^2}{\sigma_{a_k}^2}
            + \sum_k \frac{\gamma_k^2}{\sigma_{\gamma_k}^2}
        \Bigg] \,,
\end{align}
where $N^\text{obs}_{\alpha,ij} \equiv N^\text{obs}_\alpha(E_i, \theta_{\text{oa},j})$ is the ``observed'' number of events in the $i$-th energy bin and $j$-th angular bin (i.e., the number of events in the Asimov data set), and $N^\text{pred}_{\alpha,ij}(\vec{r}; \vec{a}) = N^\text{pred}(E_i, \theta_{\text{oa},j}; \vec{r}; \vec{a})$ is the number of predicted events in the same bin. It depends on the coefficients from \cref{eq:flux-superposition} -- our fit parameters -- and the vector of nuisance parameters, $\vec{a} \equiv (a_{\text{QE}},\allowbreak a_{\text{RES}},\allowbreak a_{\text{DIS}},\allowbreak a_{\text{MEC}},\allowbreak \gamma_{\text{QE}},\allowbreak \gamma_{\text{RES}},\allowbreak \gamma_{\text{DIS}},\allowbreak \gamma_{\text{MEC}})$.
Additionally, we include a bin-to-bin uncorrelated error of 1\%. We are using 500 bins linear in energy between 0.125-125 GeV corresponding to the format of the flux files from \cite{fieldsfluxes}. 

\begin{table}
    \centering
    \addtolength{\tabcolsep}{4pt}
    \begin{tabular}{c|c@{\ \ }c|c@{\ \ }c}
        \toprule
        \textbf{Parameter}
           & \multicolumn{2}{c|}{\textbf{$\nu$-dominated beam}}
           & \multicolumn{2}{c}{\textbf{$\bar\nu$-dominated beam}} \\
        \cmidrule(lr){2-3} \cmidrule(l){4-5}
           & equal time & 50\% on-axis & equal time & 50\% on-axis \\
        \midrule
        $r_{K_L}$ & $10.8\%$ & $12.1\%$ & $7.5\%$ & $8.7\%$ \\
        $r_{K^+}$ & $1.9\%$ & $1.9\%$ & $2.0\%$ & $2.2\%$ \\
        $r_{K^-}$ & $3.8\%$ & $4.5\%$ & $2.3\%$ & $2.6\%$ \\
        $r_{\pi^+}$ & $1.8\%$ & $1.8\%$ & $2.0\%$ & $2.1\%$ \\
        $r_{\pi^-}$ & $4.1\%$ & $4.4\%$ & $1.7\%$ & $1.7\%$ \\
        $a_{\text{QE}}$ & $2.4\%$ & $2.4\%$ & $2.4\%$ & $2.4\%$ \\
        $a_{\text{RES}}$ & $2.4\%$ & $2.4\%$ & $2.4\%$ & $2.4\%$ \\
        $a_{\text{DIS}}$ & $2.9\%$ & $2.9\%$ & $3.0\%$ & $3.1\%$ \\
        $a_{\text{MEC}}$ & $3.0\%$ & $3.0\%$ & $3.0\%$ & $3.0\%$ \\  
        $\gamma_{\text{QE}}$ & $1.2\%$ & $1.2\%$ & $1.2\%$ & $1.2\%$ \\
        $\gamma_{\text{RES}}$ & $1.6\%$ & $1.7\%$ & $1.5\%$ & $1.6\%$ \\
        $\gamma_{\text{DIS}}$ & $0.8\%$ & $0.8\%$ & $0.9\%$ & $0.8\%$ \\
        $\gamma_{\text{MEC}}$ & $3.5\%$ & $3.4\%$ & $3.5\%$ & $3.5\%$ \\ 
        \bottomrule
        \end{tabular}
    \addtolength{\tabcolsep}{-4pt}
    \caption{Projected fractional uncertainties on our fit parameters $r_p$ (the normalization of the various flux components) and the nuisance parameters $a_k$ and $\gamma_k$ (parameterizing the cross section uncertainties) from a fit to \SI{6.6e21}{pot} of DUNE near detector data. We show results for a neutrino-dominated beam (``forward horn current'') and for an anti-neutrino-dominated beam (``reverse horn current''), and for two different running strategies (equal data-taking time at each of the seven detector positions, or 50\% of the data-taking time spent in the on-axis positions and the rest equally distributed among six off-axis positions.)}
    \label{tab:uncertainties}
\end{table}

\begin{figure}
    \centering
    \includegraphics[width=0.8\textwidth]{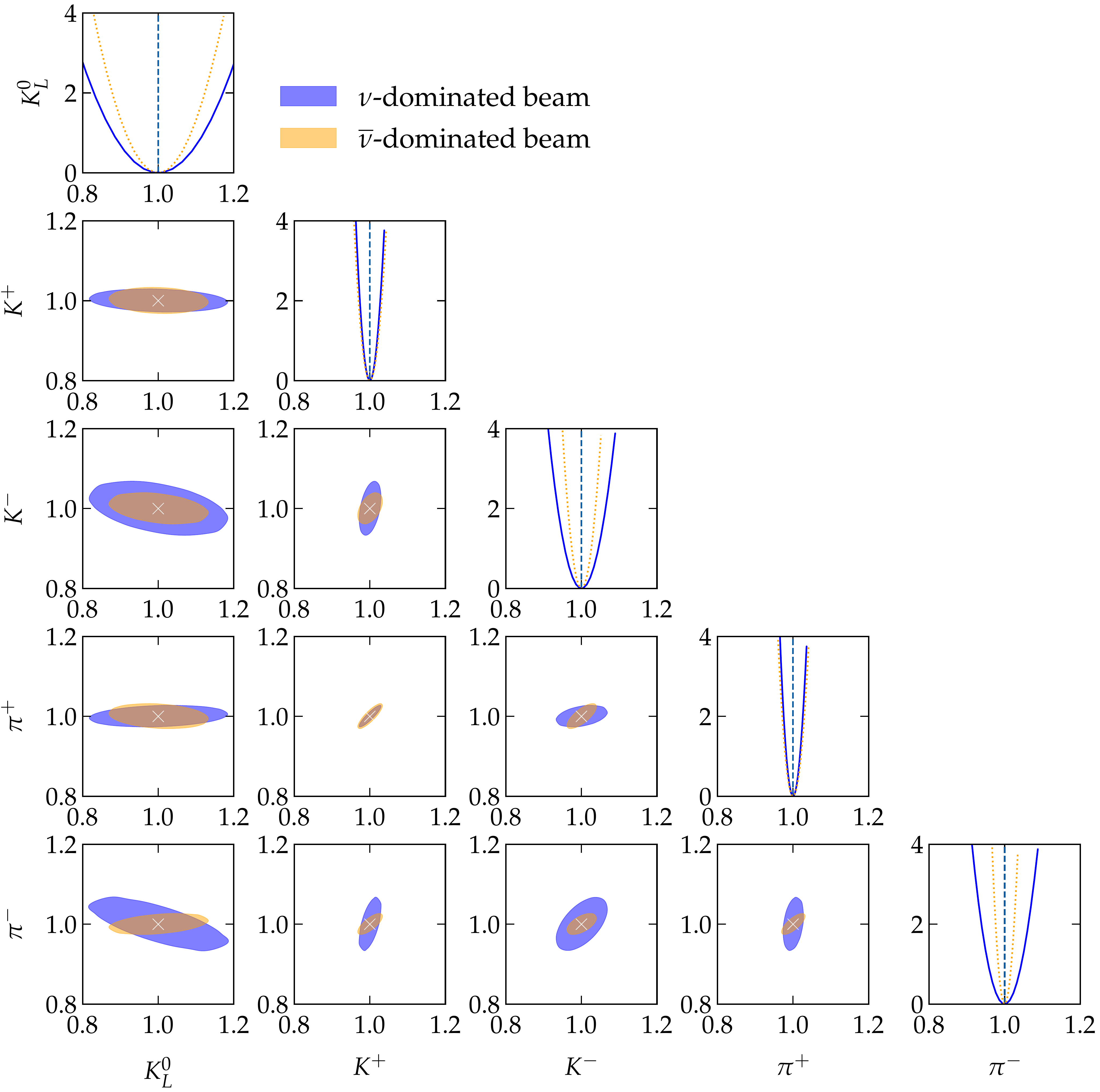}
    \caption{Projected constraints on the individual components of the DUNE neutrino flux from a fit to PRISM data in forward horn current (neutrino mode) mode (orange) and reverse horn current (antineutrino mode) mode (blue), assuming half of the total \SI{6.6e21}{pot} are collected in the on-axis position, while the rest are equally split over six different off-axis positions (\SI{6.3}{m}, \SI{12.6}{m}, \SI{18.9}{m}, \SI{24.3}{m}, \SI{30.6}{m}, \SI{36.0}{m}, corresponding to off-axis angles out to 0.063 radians).}
    \label{fig:uncertainties}
\end{figure}

\begin{figure}
    \centering
    \includegraphics[width=0.5\linewidth]{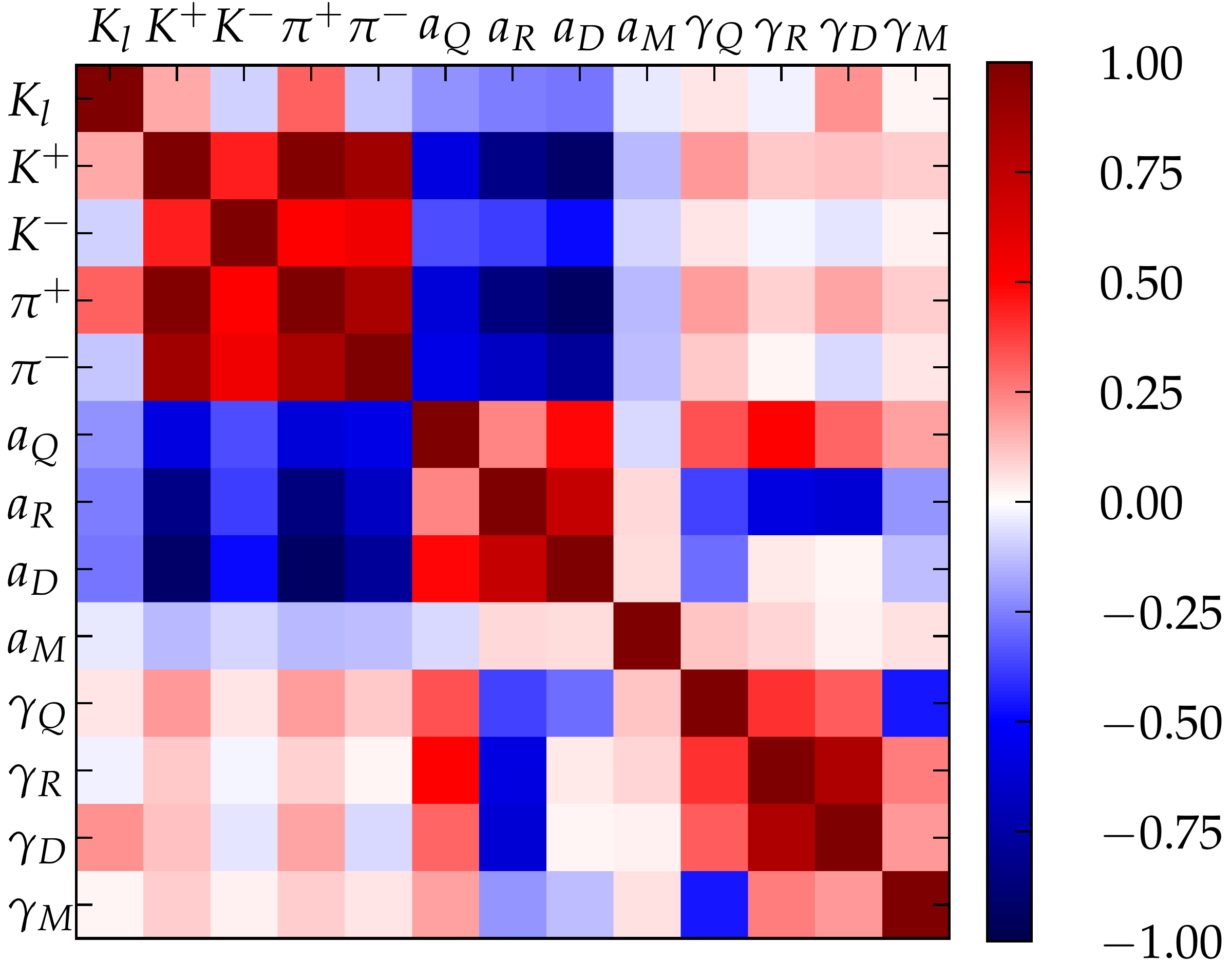}
    \caption{The correlation matrix obtained from the fit of our neutrino flux model to simulated PRISM data for the $50\%$ on-axis running strategy with a $\nu$-dominated beam.}
    \label{fig:cor_mat}
\end{figure}

The results of our fit, presented in \cref{tab:uncertainties} and \cref{fig:uncertainties}, suggest that, in our relatively simple flux model, most flux components can be constrained to the few per cent level. This should be compared to the a priori hadroproduction uncertainties of order 10\% \cite{DUNE:2020ypp}. The components from $\pi^+$ and $\pi^-$ decay, which dominate the flux in the energy range most relevant to oscillation analyses, are even constrained to better than 2\% accuracy. The weakest constraint is the one on neutrinos from $K_L$ decay. Being neutral, $K_L$ are not focused by the horns, and all semileptonic $K_L$ decay modes have at least three-body final states, so that $K_L$ make only a very small contribution to the total neutrino flux (see also \cref{fig:fluxes}).

Of course the results presented here hinge on the validity of the simplifying assumptions used in our flux model and our fit. It is clear that further improvements to the modelling of systematic uncertainties, going beyond just normalization and tilt errors, are desirable to arrive at more robust results once experiments become severely systematics-limited. For example, it would be desirable to allow for more complex variations of the spectral shape or to include the dependence of the detector efficiencies on the cross section nuisance parameters. In spite of this word of caution, our results do illustrate that off-axis near detector data is extremely powerful in improving the flux model.

Overall, the constraints we find are largely independent of the running strategy (equal time vs.\ 50\% on-axis). The $\pi^+$ flux is more strongly constrained than the one from $\pi^-$ in a neutrino-dominated beam, while for the anti-neutrino beam the situation is opposite. This is not surprising, given that the neutrino-dominated beam is created by focusing $\pi^+$ and defocusing $\pi^-$, so that neutrinos from $\pi^+$ decay account for most of the statistics in the detector. It is interesting to observe, though, that in anti-neutrino mode constraints on the different flux components are more similar than in neutrino mode, where some components are constrained very tightly, while others are only loosely bounded. The reason is that in anti-neutrino mode, the relative contributions of the different flux components to the event rate are more similar. This is because in the relevant energy range, neutrino cross sections are larger than anti-neutrino cross sections by about a factor of 3, so even though the beam is anti-neutrino dominated, neutrinos still contribute non-negligibly, leading to a wrong-sign lepton contamination. Therefore, knowing the flux more precisely can help in understanding backgrounds for oscillation analyses. 

In \cref{fig:uncertainties,fig:cor_mat} we observe significant correlations between some flux components, in particular pions and kaons. For instance, a bias of the pion contribution is easier to hide when it is accompanied by a similar bias in the kaon contribution. This is because our fit is largely insensitive to changes in the overall flux normalisation, which would be absorbed into the nuisance parameters. It is, however, sensitive to changes in the neutrino spectrum. If the pion and kaon contributions are varied simultaneously, the net effect is mostly a shift in the overall normalisation. Biasing the pions, but not the kaons (or vice-versa), on the other hand, leads to an easily detectable distortion of the spectrum.

\begin{figure}
    \centering
    \includegraphics[width=1\linewidth]{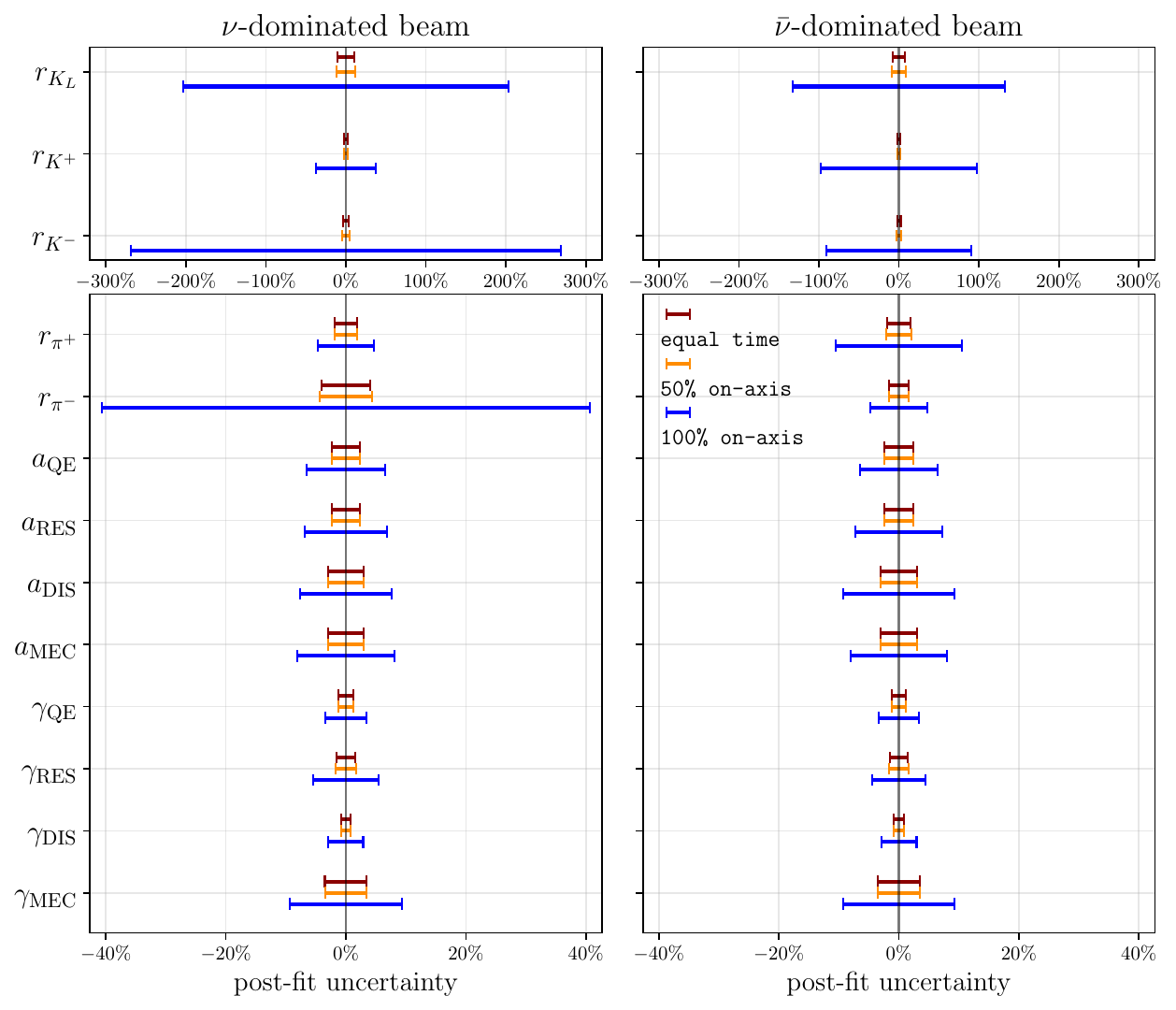}
    \caption{
Comparison of projected fractional uncertainties on the fit parameters $r_p$, the normalization of the neutrino flux components, and the nuisance parameters $a_k,~\gamma_k$ parametrizing the cross section uncertainties assuming different running strategies (equal time at each of the 7 off-axis location, 50\% and 100\% of the time on-axis in a neutrino-dominated beam (left) and anti-neutrino dominated beam (right). We have assumed $6.6\times 10^{21}$ pot of DUNE near detector data.  
    }
    \label{fig:onvsoff}
\end{figure}

Finally, we compare our results of spending equal time at the 7 off-axis location to spending 50\% or 100\% of the data-taking time on-axis in fig.~\ref{fig:onvsoff} for the neutrino-dominated and anti-neutrino-dominated beam. We find that  the uncertainty on all fit parameters is reduced by spending some time off-axis, while spending equal time on each of the off-axis location or 50\% of the time on-axis leading to similar results. In particular, the uncertainties on the kaon flux normalization in neutrino or anti-neutrino dominated beam are drastically reduced in addition to the $\pi^-$ normalization in a neutrino-dominated beam when taking off-axis data. 

\section{Impact on Oscillation Measurements}
\label{sec:osc}

We now investigate how improved modelling of the neutrino flux through LENS can benefit DUNE's main physics goals -- the precise measurement of neutrino oscillation parameters. Additional benefits could be achieved if the neutrino flux model were improved using other means, for instance improved theoretical modelling, hadroproduction experiments such as NA61/SHINE \cite{NA49-future:2006qne, NA61:2014lfx,NA61SHINE:2025aey}, or monitored or tagged neutrino beams as in nuSCOPE \cite{Acerbi:2025wzo}.

We simulate the DUNE experiment using GLoBES \cite{Huber:2004ka, Huber:2007ji}, supplemented by code developed in refs.~\cite{Kopp:2006wp, Kopp:2011qd, Kopp:2013vaa, Dentler:2018sju} and using the GLoBES implementation of DUNE from ref.~\cite{DUNE:2021cuw}. We have expanded the latter by implementing PRISM predictions in GLoBES. For each set of oscillation parameters we wish to test against the ``data'', we first determine the PRISM coefficients $c_j$ from simulations based on a (possibly biased) flux model in which the normalizations of the individual meson fluxes, $r_p$, may deviate from their nominal values (while keeping the total flux constant). We then use the resulting $c_j$ to derive an oscillated far detector flux prediction from simulated near detector ``data'' at 37 different off-axis positions (plus the on-axis position.) Simulated near detector data is based on the nominal flux model QGSP\_BERT (version glbne-v3r5p10-22-gf42572f) from ref.~\cite{fieldsfluxes}. The far detector prediction is finally compared to simulated far detector ``data'', again generated using the nominal flux model.

\begin{figure}
    \centering
    \includegraphics[width=0.75\textwidth]{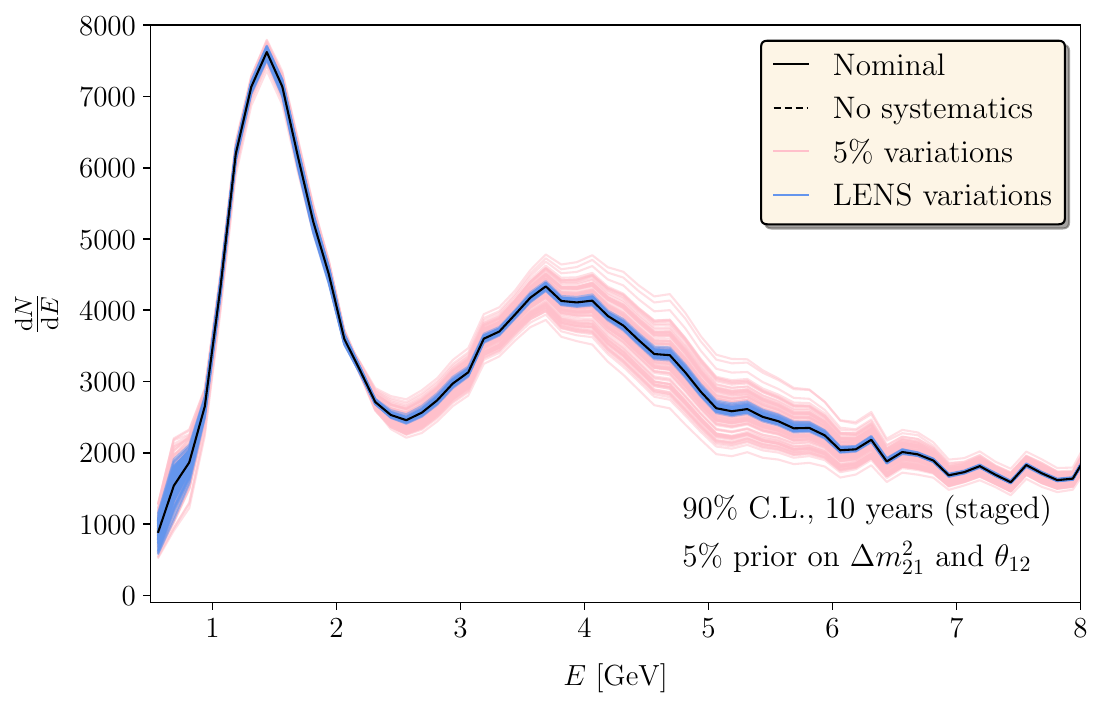}
    \caption{Impact of the flux uncertainty on the PRISM prediction of the DUNE far detector event spectrum with and without using the LENS procedure to constrain the flux model. All curves were obtained by applying \cref{eq:PRISM-prediction} to near detector data generated using the nominal flux model. The PRISM coefficients $c_j$ in \cref{eq:PRISM-1,eq:PRISM-coefficients,eq:PRISM-prediction}, on the other hand, were derived from biased flux models in which the relative importance of individual flux components (i.e., the $r_p$ coefficients in \cref{eq:flux-superposition}) were allowed to vary either by 5\% (red curves) or within the tighter constraints imposed by our LENS fit to on-axis and off-axis near detector data, see \cref{tab:uncertainties} (2nd and 4th column). In each case, we show 100 random realizations. The black curve shows for comparison the far detector spectrum based on the nominal flux model. We see that the spread in possible outcomes is substantially reduced when the LENS constraint is included, demonstrating how LENS helps making DUNE oscillation results more fiducial.}
    \label{fig:spectrum}
\end{figure}

\Cref{fig:spectrum} shows the impact of the LENS procedure on the far detector event spectrum, predicted by PRISM based on nominal near detector data. To obtain the blue curves, the $r_p$ were allowed to vary within the constraints imposed by the LENS fit from \cref{sec:flux-constraints}. More precisely, we have drawn the $r_p$ from a multivariate normal distribution parameterised by the covariance matrix from the LENS fit. We show in red how the spread of PRISM predictions increases when the $r_p$ are instead drawn from normal distributions with a 5\% width. The value 5\% is meant to represent a future flux model based on improved data from hadroproduction experiments and other external measurements, as well as theory improvements.\footnote{We make somewhat optimistic assumptions on these improvements here, as we assume that also our predictions for the improvements due to LENS, summarized in \cref{tab:uncertainties}, may be on the optimistic side given our relatively simple flux model and systematics treatment. Our main conclusion, namely that PRISM predictions will benefit significantly from improved flux modelling, remains unaffected by these caveats.} The black curve finally shows the far detector prediction for the nominal flux model, where $r_p = 1$ for all flux components. For the overall normalization, we assume 10~years of data taking (\SI{7}{yrs} at \SI{1.2}{MW} beam power, \SI{3}{yrs} at \SI{2.4}{MW}, corresponding to \SI{1.6e23}{pot}) with \SI{40}{kt} fiducial detector mass.\footnote{We adopt here the convention that one ``Fermilab-year'' corresponds to \SI{1e7}{sec}, taking into account a typical $\sim 30\%$ average duty cycle of the Fermilab accelerator complex.} It is evident from \cref{fig:spectrum} that better constraints on the flux model, like the ones that can be obtained from the LENS fit or from external measurements and calculations, can substantially reduce the uncertainty in the far detector flux predictions.

\begin{figure}
    \centering
    \includegraphics[width=0.95\textwidth]{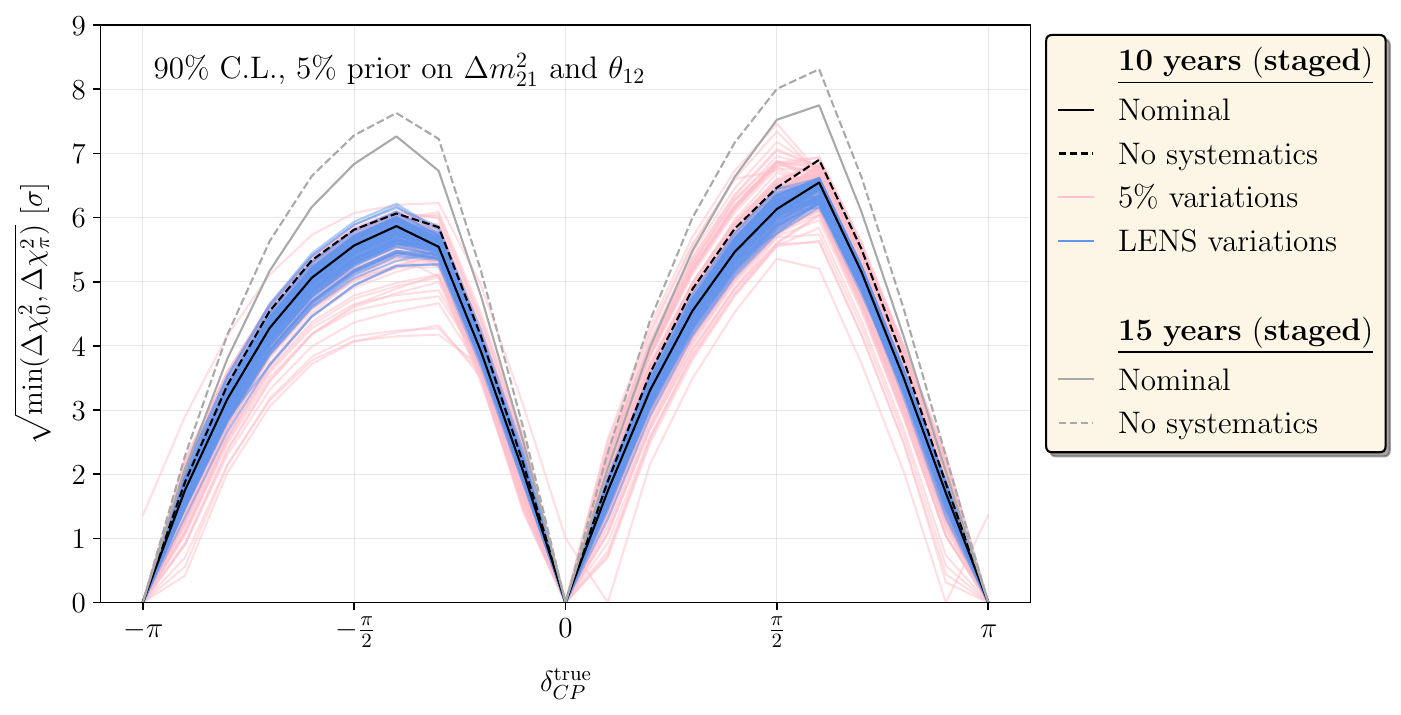}
    \caption{The ``McDonald's plot'' showing the impact of neutrino flux uncertainties on DUNE’s sensitivity to leptonic CP violation with \SI{1.6e23}{pot} at \SI{40}{kt} fiducial detector mass (7~years of running at \SI{1.2}{MW} beam power + 3~years at \SI{2.4}{MW} = \SI{624}{kt\,MW\,yrs}, equally split between FHC and RHC). For each value of the true $\delta_{CP}$, we show the significance at which the CP-conserving hypothesis ($\delta_{CP} \in \{ 0, \pi \}$) can be excluded. Similar to \cref{fig:spectrum}, all curves were obtained using PRISM to predict the oscillated far detector fluxes based on near detector data. While the near and far detector data were simulated using the nominal flux model, the PRISM coefficients $c_j$ were derived based on biased flux models (red: 5\% variations, blue: variations allowed from \cref{tab:uncertainties}). Note that the systematic uncertainties assumed in the oscillation fit are the same for all curves, even though LENS would allow us to reduce them, boosting the sensitivity. The black dotted curve shows the hypothetical sensitivity without systematic uncertainties. In grey, we illustrate how the impact of systematic uncertainties grows with larger exposure (15~years, corresponding to \SI{1104}{kt\,MW\,yrs}).}
    \label{fig:osci_results1}
\end{figure}

As a first oscillation result, we show in \cref{fig:osci_results1} DUNE's projected sensitivity to leptonic CP violation as a function of the true CP phase, $\delta_{CP}^\text{true}$. (Due to its characteristic shape, this type of plot is sometimes called the McDonald's plot.) For each value of $\delta_{CP}^\text{true}$, the plot shows the significance at which this $\delta_{CP}^\text{true}$ can be distinguished from the hypothesis of CP conservation ($\delta_{CP} \in \{ 0, \pi \}$). All other oscillation parameters as well as the matter density along the neutrino trajectory have been marginalised over. For the solar parameters $\theta_{12}$ and $\Delta m_{21}^2$, to which DUNE does not have good sensitivity \cite{Denton:2023zwa}, as well as for the matter density, external 5\% priors have been implemented \cite{Esteban:2024eli}. Each curve corresponds to a different flux model, that is, different values of the coefficients $r_p$ in \cref{eq:flux-superposition} have been used to determine the PRISM coefficients $c_j$ in \cref{eq:PRISM-1,eq:PRISM-coefficients,eq:PRISM-prediction}. In contrast, the near detector ``data'' which enters \cref{eq:PRISM-prediction} is always based on the nominal flux model. The meaning of the coloured curves in \cref{fig:osci_results1} is the same as in \cref{fig:spectrum} above. We see that first constraining the flux model has significant benefits for the PRISM prediction. LENS ensures that the flux model employed to compute the PRISM coefficients $c_j$ is close to the model realised in Nature. If this is not done, the sensitivity may be overestimated or underestimated by up to $1\sigma$.

In principle, improving our understanding of the neutrino flux with LENS could justify reducing the systematic uncertainties implemented in the oscillation fit, thereby actually improving the sensitivity to the oscillation parameters (rather than just making the exclusion curves more reliable). We refrain from showing this improvement here as doing so would require us to disentangle the residual flux uncertainty from any detector-related uncertainties in the systematics model of Ref.~\cite{DUNE:2021cuw}. Moreover, as explained in \cref{sec:flux-constraints}, we consider our estimates for the accuracy of LENS as at best indicative. Constructing a reliable systematics model based on LENS will require a more detailed study than the one presented here. Instead, we show in \cref{fig:osci_results1} the expected sensitivity to CP violation in the absence of \emph{any} systematic uncertainties (black dotted curve). This curve is meant to illustrate qualitatively the maximum improvement in sensitivity achievable through reduction of systematic uncertainties. In grey, we illustrate how the impact of systematic uncertainties, and the potential benefit of reducing them, grows at a larger exposure.

\begin{figure}
    \centering
    \includegraphics[width=0.75\textwidth]{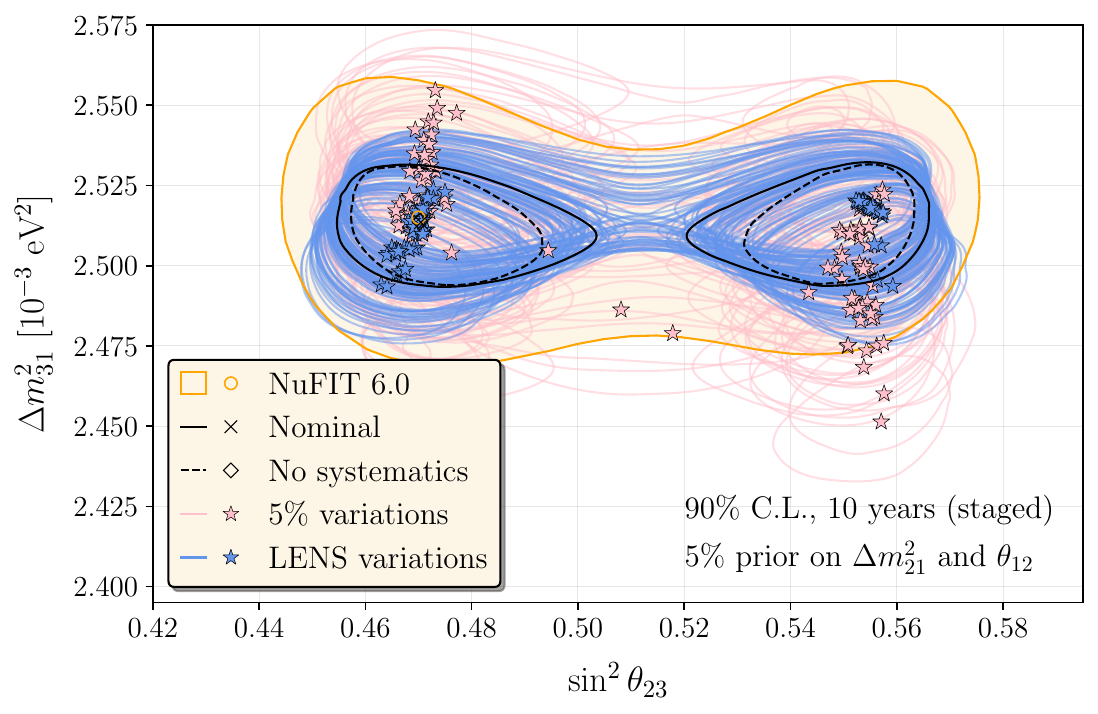} \\
    \includegraphics[width=0.75\textwidth]{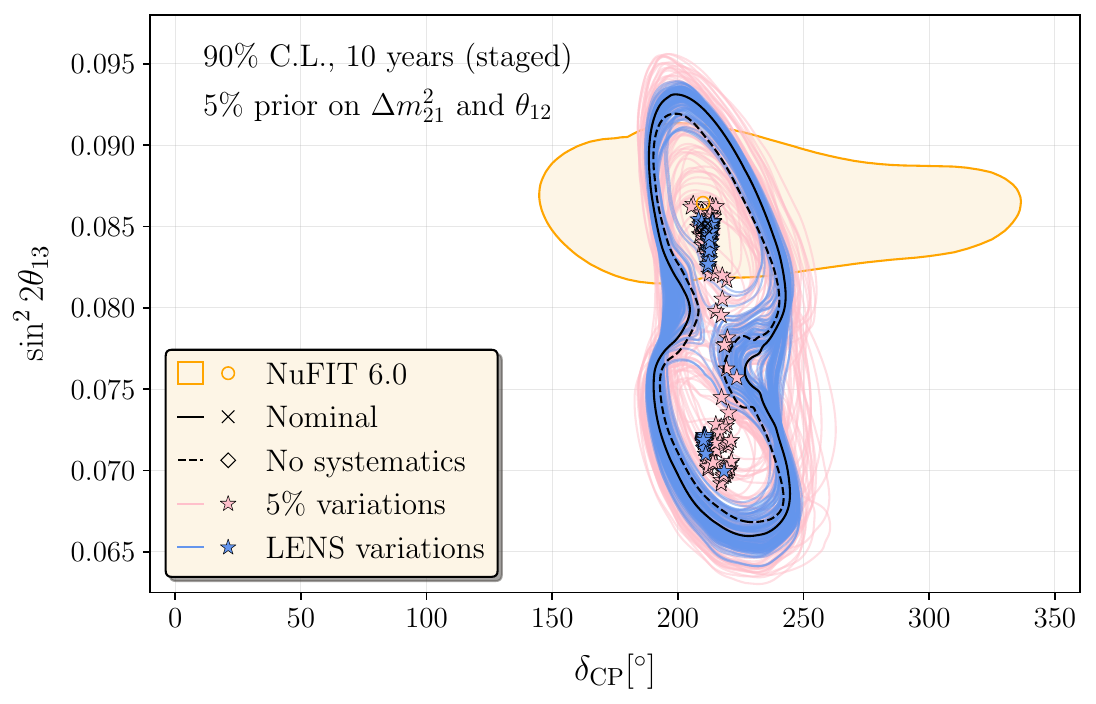}
    \caption{Impact of LENS on the measurement of the oscillation parameters $(\theta_{13}, \delta_{CP})$ (top, the ``superhero plot'') and $(\theta_{23}, \Delta m_{31}^2)$ (bottom, the ``peanut plot'') with \SI{1.6e23}{pot} at \SI{40}{kt} fiducial detector mass. As in \cref{fig:osci_results1}, red and blue contours correspond to neutrino flux models with and without the LENS constraint; black solid curves corresponds to the nominal model and black dotted ones to an analysis with only statistical uncertainties. In all cases, the far detector prediction is based on PRISM, with the superposition coefficients $c_j$ determined from biased flux models, while the ``data'' has been generated using the nominal model. For comparison, we  show in bright orange the current constraints on the oscillation parameters from NuFit~6.0 \cite{Esteban:2024eli}.}
    \label{fig:osci_results2}
\end{figure}

The observations from \cref{fig:osci_results1} are confirmed by \cref{fig:osci_results2}, which shows two-dimensional projections of DUNE's anticipated sensitivity contours in $\theta_{13}$, $\delta_{CP}$, $\theta_{23}$, and $\Delta m_{31}^2$. Due to their characteristic shape, we call these plots the ``superhero plot'' and the ``peanut plot''. Again, we have marginalised over all oscillation parameters that do not appear on the axes, and we have taken into account an external prior on the solar parameters. As in \cref{fig:osci_results1} above, we see that constraining neutrino fluxes with LENS significantly reduces the spread in possible outcomes, thereby making results more robust and more reliable. Again, LENS would also allow for an oscillation fit with reduced systematic errors, which would improve the sensitivity to somewhere between the black solid and black dotted contours.

\section{Conclusions}
\label{sec:conclusions}

We have demonstrated how improved flux modelling can enhance the accuracy of precision neutrino oscillation measurements in accelerator-based experiments such as DUNE and Hyper-Kamiokande. While far detector predictions based on on-axis and off-axis near detector measurements (e.g., DUNE-PRISM or IWCD in Hyper-Kamiokande) already go a long way in reducing systematic uncertainties, we demonstrate in \cref{fig:osci_results1,fig:osci_results2} how our imperfect understanding of neutrino fluxes still limits the accuracy of oscillation measurements.

We have therefore investigated the LENS (Lateral Extraction of Neutrino Spectra) method, which consists of first performing a near detector-only fit of a parametric neutrino flux model, leveraging on-axis and off-axis measurements to constrain the relative contributions of individual flux components, in particular, neutrino fluxes from different parent meson species.

We have then studied how uncertainties in the flux model affect the PRISM prediction of the oscillated fluxes at the far detector and thereby ultimately the oscillation fit. We have found that, with the LENS constraint included, possible biases in DUNE oscillation constraints are substantially reduced. Complementary improvements could come from future hadroproduction experiments or from an instrumented decay region as in nuSCOPE, further boosting the accuracy of oscillation analyses via improvements on the neutrino cross sections.

In conclusion, our work shows how long-baseline oscillation experiments can optimally benefit from their near detectors. As a corollary, it also implies that highly capable near detectors are crucial for the success of these experiments. In the systematics-dominated regime in which DUNE and Hyper-Kamiokande will ultimately operate, even modest improvements in near detector performance can lead to substantially improved oscillation results.

\section*{Acknowledgments}
It is a pleasure to thank Laura Fields for many useful conversations concerning the simulation of the DUNE fluxes and for making her results available in machine-readable form.
We would also like to thank Alfons Weber, Ioana Caracas and Mike Wilking for sharing their expertise on DUNE-PRISM. We thank Alfons Weber, Lukas Koch, and Alex Wilkinson for useful comments and discussions on the preprint.
JG and GAP acknowledge useful discussions at the NEAT workshop held at Colorado State University in May 2025. JG acknowledges support by the U.S.\ Department of Energy Office of Science under award number DE-SC0025448. GAP's work has been supported by the Cluster of Excellence ``Precision Physics, Fundamental Interactions, and Structure of Matter'' (PRISMA+ EXC 2118/1) funded by the German Research Foundation (DFG).

\bibliographystyle{JHEP}
\bibliography{main}

\end{document}